\documentclass[11pt]{article}
\ifdefined\pdftexversion\pdfoutput=1\fi

\usepackage[margin=1in]{geometry}
\usepackage[T1]{fontenc}
\usepackage{lmodern}
\usepackage{microtype}
\usepackage{graphicx}
\usepackage{booktabs}
\usepackage{tabularx}
\usepackage{array}
\usepackage{amsmath}
\usepackage{amssymb}
\usepackage{xcolor}
\usepackage[numbers,sort&compress]{natbib}
\usepackage{enumitem}
\usepackage[colorlinks=true,linkcolor=black,citecolor=black,urlcolor=blue]{hyperref}
\usepackage{titlesec}

\titleformat{\section}{\normalfont\large\bfseries}{\thesection}{0.6em}{}
\titleformat{\subsection}{\normalfont\normalsize\bfseries}{\thesubsection}{0.6em}{}

\newcommand{\code}[1]{\texttt{\small #1}}
\newcolumntype{L}[1]{>{\raggedright\arraybackslash}p{#1}}

\title{\bfseries Elastic KV Cache for LLM Serving:\\
A Working Reclamation Mechanism, and Why Chunked Prefill Already Closes the Gap}
\author{Sathishkumar Sivashanmugam\\ \small Amazon Web Services}
\date{\today}

\usepackage{caption}
\begin{document}
\maketitle

\begin{abstract}
An LLM serving engine sizes its key-value (KV) cache once, at startup. To do that it
permanently sets aside a reserve for the worst-case prefill activation. During
decode-dominant phases that reserve sits idle, yet it cannot be handed to the KV pool
because it is exactly the memory a large prefill needs. We ask whether this reserve is a
reclaimable resource, so we built a mechanism to test it. Our elastic KV cache lends the reserve
to the KV pool during decode phases and returns it before prefill, driven by the
scheduler's one-step-ahead view of the next batch. The mechanism is pure userspace on the
CUDA virtual-memory path: two physical handles mapped into one contiguous virtual range
per layer, so the attention kernel is unchanged and no driver patch is required. It
decommits the reserve in a few milliseconds and recommits it in tens of milliseconds, works
with CUDA graphs and prefix caching enabled, and never triggers an out-of-memory (OOM)
event. We show that a static commit of the same memory is unsafe. It crashes on prefill
bursts, which makes the dynamic toggle necessary to capture the reserve at all.

Having built the mechanism, we then test the premise it rests on, and report an honest
negative result. The technique only pays off if a small prefill chunk size badly hurts
prefill latency, forcing operators to choose between prefill throughput and KV capacity.
In a controlled experiment that injects long prompts into a live decode load, that penalty
turns out to be small (median time-to-first-token differs by about 1\% between chunk sizes
of 8192 and 32768 tokens). The reason is structural: prefill is compute bound, so chunking
splits the same work across steps, and decode consumes only about one token per sequence
per step, so it never starves the prefill budget. Simply lowering
\code{max\_num\_batched\_tokens} recovers more KV than the elastic controller does, at
nearly equal latency, so the controller does not beat that trivial baseline. We further
show the reserve dilutes under tensor parallelism, shrinking from 16\% of KV at TP1 to as
little as 2.7\% at TP4. Large models need high tensor parallelism, so this regime does not
help either. We conclude with a
precise statement of when reclaiming the reserve could still help, and we release the
mechanism as a reusable userspace elastic-VMM allocator.
\end{abstract}

\section{Introduction}
Modern LLM serving engines such as vLLM~\cite{vllm} manage GPU memory by carving the
device into a few fixed regions at startup: model weights, a CUDA-graph pool, a small
activation working set, and everything left over becomes the KV cache. The KV cache is the
scarce resource. It bounds both the number of concurrent requests and the maximum context
length, so operators want it as large as possible.

The size of the KV cache is not a free parameter. Before it is sized, the engine profiles
a worst-case forward pass and sets aside enough memory to run the largest prefill it will
ever schedule. The larger the prefill chunk, controlled by
\code{max\_num\_batched\_tokens}, the larger this activation reserve, and the smaller the
KV cache. This sets up an apparent tradeoff. A large chunk gives fast prefill but a small
KV cache; a small chunk gives a large KV cache but, one would assume, slow prefill. The
reserve is provisioned for the peak and is idle whenever the engine is not running a large
prefill, which for many workloads is most of the time.

This paper studies whether that idle reserve can be reclaimed. We treat the
prefill-activation reserve as a resource that can be lent to the KV cache during
decode-dominant phases and returned before the next large prefill, and we build a working
mechanism to do exactly that. The design question is entirely about the memory substrate:
the KV cache in vLLM is one contiguous tensor per layer, and the attention kernel indexes
it by block id, so any scheme that grows and shrinks it must preserve a single contiguous
virtual address range per layer or else touch the kernel. We solve this on the CUDA
virtual-memory (VMM) API by reserving one virtual range per layer and backing it with two
physical handles, a base handle and an elastic handle, mapping and unmapping only the
elastic sub-range. The attention kernel never changes, no driver is patched, and the toggle
costs a few milliseconds for the full 3~GiB reserve.

We report two kinds of result. First, the mechanism works and is necessary: a static commit
of the reserve crashes on prefill bursts, while the dynamic toggle runs the same bursts
clean, and on a low-tensor-parallel, long-context configuration the toggle reclaims about
18\% more KV capacity and about 10\% more decode throughput at zero OOM. Second, the premise
does not hold on the workloads that matter. The prefill
penalty of a small chunk size is small because prefill is compute bound and decode is cheap
per step, so vLLM's chunked prefill already dissolves the tradeoff the technique was built to
exploit. Lowering the chunk size is a better lever than an elastic cache. We show this with a
direct time-to-first-token (TTFT) measurement, and we show that tensor parallelism removes the
only regime where the reserve is large.

Our contributions are:
\begin{itemize}
\item A userspace elastic KV mechanism on the CUDA VMM path that grows and shrinks the KV
cache without changing the attention kernel or patching the driver, and toggles the full
reserve in milliseconds (Section~\ref{sec:mech}).
\item A necessity result: static reclamation of the reserve is unsafe and OOMs on prefill
bursts, so only dynamic toggling can capture it (Section~\ref{sec:necessity}).
\item An honest negative result: a controlled TTFT-under-load experiment shows the prefill
penalty of a small chunk size is about 1\% at the median, so simply lowering
\code{max\_num\_batched\_tokens} recovers more KV at equal latency and the elastic controller
does not beat it (Section~\ref{sec:decisive}).
\item A scope characterization, including a tensor-parallel dilution law that shrinks the
reserve from 16\% to as little as 2.7\% of KV, that tells practitioners when reclaiming the
reserve could be worthwhile and when it cannot (Section~\ref{sec:scope}).
\end{itemize}

All measurements are on a single NVIDIA A100-SXM4-40GB, Qwen2.5-7B-Instruct in float16,
vLLM 0.23.0, \code{gpu\_memory\_utilization}=0.9 and \code{max\_model\_len}=32768 unless
stated otherwise.

\section{Background and Motivation}
\label{sec:bg}
\paragraph{KV allocation in vLLM.}
vLLM allocates the KV cache as a paged block pool~\cite{vllm}. At startup it loads weights,
runs a profiling forward pass to measure the peak activation memory of the largest allowed
batch, reserves that plus a CUDA-graph pool, and turns the remainder into fixed-size KV
blocks. The block pool is a pure integer free list with no runtime allocation. Blocks are
handed to requests and returned on completion. The pool is already a slab allocator, so
there is no fragmentation to reclaim inside it. The reclaimable memory, if any, lives
in the reserves outside it.

\begin{figure}[t]\centering
\includegraphics[width=0.62\linewidth]{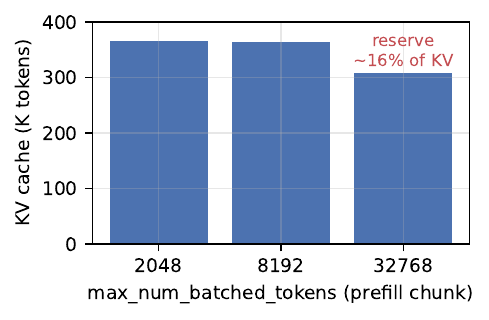}
\caption{The prefill-activation reserve. KV capacity shrinks as the prefill chunk grows,
because the engine reserves more memory for the prefill activation peak. The gap is about
16\% of KV at a 32768-token chunk (Qwen2.5-7B, TP1, util=0.9).}
\label{fig:reserve}
\end{figure}

\paragraph{The prefill-activation reserve.}
The activation reserve scales with the prefill chunk. Figure~\ref{fig:reserve} shows KV
capacity as we raise \code{max\_num\_batched\_tokens}. Going from 2048 to 32768 tokens per
step shrinks the KV cache from about 366K to about 308K tokens, because the engine now
reserves roughly 3.1~GiB for the larger prefill activation peak. That 3.1~GiB is about 16\%
of the KV cache. It is committed at startup and sits free whenever the engine is decoding
rather than running a large prefill.

\begin{figure}[t]\centering
\includegraphics[width=0.62\linewidth]{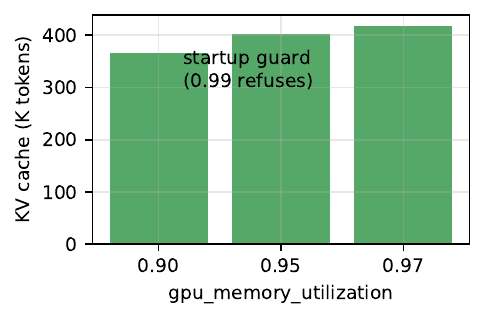}
\caption{Raising \code{gpu\_memory\_utilization} captures stranded headroom but plateaus at
the startup guard (0.99 refuses to start). This is a different pool from the activation
reserve.}
\label{fig:util}
\end{figure}

\paragraph{The utilization knob is not the same thing.}
One might ask whether raising \code{gpu\_memory\_utilization} already captures this.
Figure~\ref{fig:util} shows it captures a different pool, the stranded headroom between the
utilization target and the device. Raising the target from 0.90 to 0.97 grows KV from about
366K to 418K tokens, but the run refuses to start at 0.99 because about 0.8~GiB of CUDA
context is unreclaimable and trips the startup guard. The utilization knob captures headroom;
it does not touch the activation reserve, which is provisioned inside the utilization budget
specifically for prefill.

\section{An Elastic KV Mechanism}
\label{sec:mech}
The goal is a KV cache that can grow into the reserve during decode and shrink back before
prefill, without touching the attention kernel.

\paragraph{The contiguity constraint.}
vLLM stores the KV cache as one contiguous tensor per layer, and the FlashAttention backend
indexes that tensor by block id through a single base pointer. There is no second-tensor
parameter. Any elastic scheme therefore has two options: keep one contiguous virtual range
per layer and change only its physical backing, or split the KV into two tensors and rewrite
the kernel to read both. We built prototypes of both and they agree that the first is
strictly cheaper, because it leaves the kernel untouched. We adopt it.

\paragraph{Two handles, one virtual range.}
We reserve one virtual address range per layer sized for base plus elastic, and back it with
two physical handles through the CUDA VMM API (\code{cuMemCreate}, \code{cuMemMap},
\code{cuMemAddressReserve}). The base handle is always mapped. The elastic handle maps or
unmaps a sub-range at the top of the virtual range, aligned to the 2~MiB VMM granularity.
Because the virtual range never moves and the base sub-range stays mapped, the kernel keeps
reading one contiguous pointer per layer whether the elastic slice is present or not. We
expose this as a torch pluggable allocator so the KV tensors are backed by it transparently,
and we gate the top elastic block-id range in the block pool so those blocks are only handed
out when the elastic slice is committed.

\begin{figure}[t]\centering
\includegraphics[width=0.62\linewidth]{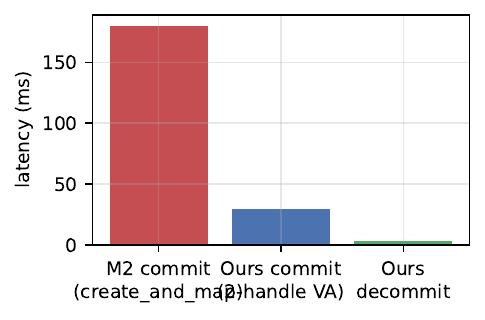}
\caption{Toggle latency for the 3~GiB reserve across 28 layers. The contiguous-VA mechanism
commits in about 30~ms and decommits in about 3~ms, versus 180~ms for a naive whole-handle
commit.}
\label{fig:latency}
\end{figure}

\paragraph{Toggle cost.}
Figure~\ref{fig:latency} shows the toggle latency for the full 3.09~GiB reserve across 28
layers. A naive whole-handle commit through \code{create\_and\_map} costs about 180~ms
because the physical allocation dominates and does not batch. Our contiguous-VA path commits
in about 30~ms and decommits in about 3~ms, since decommit only unmaps. In the live
controller with CUDA graphs on we measure decommit at 3.8~ms and recommit at 23.8~ms. These
costs are small relative to phase durations (Section~\ref{sec:scope}), and decommit, the
operation that must happen before a prefill, is the cheap direction.

\paragraph{Controller.}
The scheduler knows the next batch before it runs it, so it knows one step ahead whether the
next step is decode-only or contains a prefill. The controller commits the elastic slice
during decode-only phases and decommits it at the boundary before a prefill, with hysteresis
to avoid thrashing. Committing grows \code{num\_gpu\_blocks} and admits the top elastic
blocks. Decommitting first drains those blocks, syncs, and then unmaps. We verified that
generation is bit-identical across commit and decommit cycles, that a peak live block id
exceeding the base range is served coherently in a single kernel call, and that the scheme
works with CUDA graphs and prefix caching both enabled.

\section{Static Reclamation is Unsafe}
\label{sec:necessity}
Before asking whether dynamic toggling wins, we establish that it is necessary, meaning a
static commit of the reserve cannot safely capture it.

\begin{figure}[t]\centering
\includegraphics[width=0.62\linewidth]{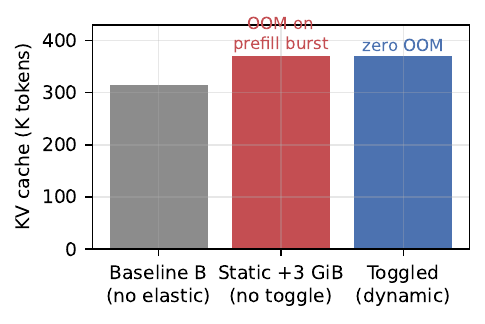}
\caption{Static reclamation of the full reserve recovers the capacity but OOMs on a prefill
burst. Only the dynamic toggle captures the full reserve safely.}
\label{fig:necessity}
\end{figure}

We statically grew the KV cache by 3~GiB, folding the whole reserve into the pool. This
recovers the full capacity gap, from about 314K to about 370K tokens. It then OOMs on the
first large prefill. A 62K-token prefill burst needs about 2.3~GiB of activation, which is
exactly the memory we lent to KV, so there is nothing left for the forward pass. The
OOM-safe static ceiling is only about 1.2~GiB, a 6\% gain, because that is all that can be
lent without starving the largest prefill. The dynamic toggle runs the same 62K-token burst
clean: it decommits the elastic slice first, in 4.4~ms, then prefills, then recommits.
Figure~\ref{fig:necessity} contrasts the two. This is the argument for the mechanism. The
full reserve is reachable only by giving it back before prefill, and only a dynamic scheme
can do that.

We note one bug worth recording: the first recommit after a large prefill OOMed because the
torch caching allocator held the freed prefill memory. Calling \code{empty\_cache()} before
\code{cuMemCreate} fixes it.

\section{The Decisive Test: Does a Small Prefill Chunk Actually Hurt?}
\label{sec:decisive}
The mechanism works and is necessary, but that is not the same as being worthwhile. The
entire value proposition rests on one assumption: that a small prefill chunk size badly hurts
prefill latency, so operators are forced to run a large chunk and pay the KV cost, which the
elastic cache then buys back. If a small chunk does not hurt, then the operator can simply
lower \code{max\_num\_batched\_tokens}, keep the reserve as KV for free, and the elastic
cache has nothing to add. This section tests that assumption directly.

\paragraph{Setup.}
We measure foreground TTFT for long prompts injected into a live decode load, which is the
one regime where a small chunk should hurt through head-of-line blocking. We run 40 background
decode sequences generating long outputs to hold a steady decode load, then inject six long
prompts of about 25K tokens each, spaced 1.5~s apart, and record the TTFT of each. We compare
three modes: A with \code{max\_num\_batched\_tokens}=8192, B with 32768, and the elastic
controller (auto) at 32768 with the 3~GiB reserve lent to KV during decode. Prefix caching and
CUDA graphs are both on. We use vLLM's asynchronous engine so the reported TTFT is the true
time from submission to first output token.

\begin{figure}[t]\centering
\includegraphics[width=0.62\linewidth]{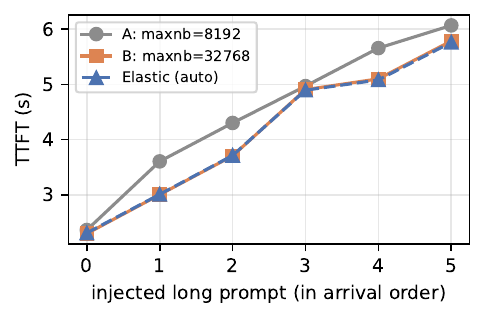}
\caption{Foreground TTFT for six long prompts injected into a live decode load. The three modes
nearly overlap; a small prefill chunk barely raises TTFT.}
\label{fig:mixed}
\end{figure}

\paragraph{Result.}
Table~\ref{tab:decisive} and Figure~\ref{fig:mixed} show the outcome. The three modes nearly
overlap. Mode B's median TTFT is about 1\% lower than mode A's, and its worst single request is
about 5\% lower. That is far from the roughly fourfold difference one would expect from a
32K-token prompt needing one step at chunk 32768 versus four steps at chunk 8192. The elastic
controller matches B's TTFT, as it must since it uses the same chunk size, and toggles the
reserve correctly, but its maximum KV of about 364K tokens is still below mode A's 375K.

\begin{table}[t]\centering
\caption{Foreground long-prompt TTFT under a live decode load, and KV capacity. Mode A (small
chunk) wins on both axes. It has the most KV, and its TTFT is within about 1\% of the others.}
\label{tab:decisive}
\small
\begin{tabular}{lccc}
\toprule
Mode & TTFT median (s) & TTFT max (s) & KV capacity (K tok) \\
\midrule
A: chunk 8192            & 4.97 & 6.07 & \textbf{375} \\
B: chunk 32768           & 4.91 & 5.79 & 314 \\
Elastic (auto)           & 4.90 & 5.76 & 364 \\
\bottomrule
\end{tabular}
\end{table}

\begin{figure}[t]\centering
\includegraphics[width=0.62\linewidth]{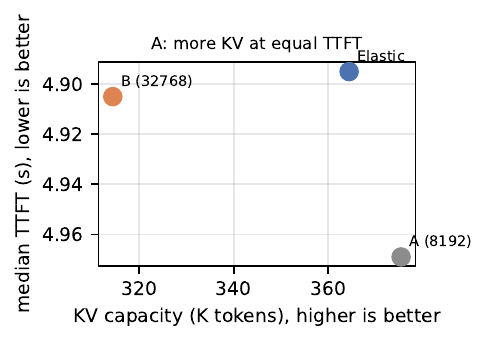}
\caption{The KV-versus-latency frontier. Mode A (small chunk) has the most KV at about the
same median TTFT, so it wins on both axes and the elastic cache has little to buy back.}
\label{fig:pareto}
\end{figure}

\paragraph{Why the penalty is small.}
The result is structural, not an artifact of the setup. Prefill is compute bound at these
sizes, so splitting a 32K-token prompt into four 8K chunks costs about the same total
work as one 32K step. The chunk size changes the granularity, not the total FLOPs. Decode
is also cheap per step. Each running sequence contributes about one token per step, so even
at the 280-sequence maximum, decode adds only about 280 tokens against a chunk budget of
8192, which is about 3\%. Decode therefore never meaningfully starves the prefill budget,
and the head-of-line effect a small chunk is supposed to cause does not appear.
Figure~\ref{fig:pareto} shows the consequence. Mode A has the most KV at the same latency,
so it wins on both axes that matter. The elastic controller can at best recover mode B's
roughly 1\% median latency edge while staying below mode A on capacity, which is not a win.

\paragraph{Robustness.}
Harsher regimes strengthen mode A rather than the elastic cache. Decode cannot consume the
prefill budget by construction, so raising the decode load does not create the starvation the
technique needs. Saturating the KV cache favors mode A, which has the larger pool and preempts
later. A prefill-throughput-bound burst processes the same total prefill work at either chunk
size, so aggregate throughput is chunk-size independent. And tensor parallelism, examined next,
shrinks the reserve. We could not find a regime in which the elastic controller beats simply
lowering the chunk size.

\section{Scope: When Could This Help?}
\label{sec:scope}
The negative result is workload specific, so we state precisely where the reserve is large and
where it is not.

\begin{figure}[t]\centering
\includegraphics[width=0.62\linewidth]{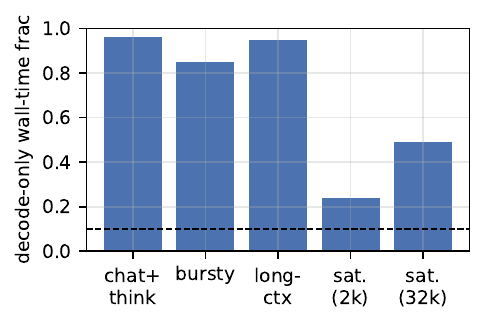}
\caption{Decode-only fraction of wall time by regime. Windows far exceed the recommit cost
except under sustained saturation, and the larger chunk has more availability.}
\label{fig:phase}
\end{figure}

\paragraph{Phase availability is not the problem.}
One might worry the gain lives only in decode-only windows that are too rare. We instrumented
the scheduler per step and measured the decode-only fraction of wall time by regime
(Figure~\ref{fig:phase}). It is 95 to 97\% for chat with think time, 72 to 97\% for bursty
traffic, 90 to 99.8\% for long context, and 24\% (chunk 2048) to 49\% (chunk 32768) under
sustained saturation. The larger chunk has the larger reserve and gain, and it also has more
decode-only availability. Phase availability is not what kills the technique.

\begin{figure}[t]\centering
\includegraphics[width=0.62\linewidth]{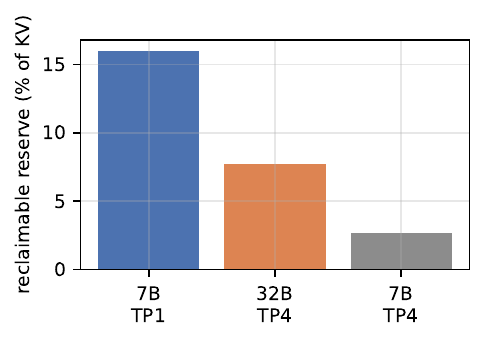}
\caption{The reserve dilutes under tensor parallelism, from 16\% of KV at TP1 to as little as
2.7\% at TP4. Big models need high TP, so their reserve is small.}
\label{fig:tp}
\end{figure}

\paragraph{Tensor parallelism dilutes the reserve.}
The reserve is a per-rank activation cost, and tensor parallelism shards it across ranks while
the per-rank KV stays large, so the reserve shrinks as a fraction of KV. Figure~\ref{fig:tp}
shows the dilution: the reserve is 16\% of KV at 7B TP1, 7.7\% at 32B TP4, and 2.7\% at 7B TP4.
Larger models help slightly at fixed TP (2.7\% to 7.7\%), but they require high TP to fit, and
high TP dominates. Big models are served with tensor parallelism, and under tensor parallelism
the reserve is small. So large models do not help either.

\paragraph{Where it could still help.}
The reserve is largest, roughly 10 to 16\% of KV, exactly when tensor parallelism is low, the
model is small to mid-sized, the context is long, and the KV cache is scarce. But that is also
the regime where lowering the chunk size is cheapest, because a small model prefills fast, so the
same negative argument applies. We did not find a configuration where the elastic cache is the
right tool. If one exists, it would need a workload where a small chunk throttles
prefill throughput and KV is scarce at the same time, which we could not construct on this
hardware.

\section{Comparison with Alternatives}
We compare against the natural alternatives for reclaiming KV memory.

\paragraph{The utilization knob (util=0.97).}
Running the default allocator at util=0.97 OOMs on the 48-prompt prefill burst due to activation
plus fragmentation, and survives only with \code{expandable\_segments} enabled, and then only by
0.04 to 0.15~GiB of margin. It is not robust. It also captures headroom, not the activation
reserve, as shown in Section~\ref{sec:bg}.

\paragraph{vAttention.}
vAttention~\cite{vattention} manages KV contiguity through demand paging on a replaced NVIDIA UVM
driver with sub-2~MiB pages. It requires patching the driver and pins specific torch and CUDA
versions, which we could not run on our stack, so we compare analytically. vAttention keeps
contiguity within a fixed KV budget. It does not make the budget elastic against the activation
reserve, and it needs a driver patch, which our mechanism avoids.

\paragraph{Jenga.}
Jenga~\cite{jenga} sizes KV blocks for heterogeneous and hybrid attention and is already upstream
in vLLM 0.23.0 as the hybrid KV coordinator. For a single-group model like Qwen it is a no-op and
behaves as stock. It sizes blocks within a fixed KV region and is orthogonal to reclaiming the
reserve, so the two could compose.

None of these reclaim the activation reserve, but as our results show, that reserve is not worth
reclaiming on the workloads we tested, because chunked prefill already makes a small chunk cheap.

\section{Related Work}
PagedAttention and vLLM~\cite{vllm} introduced the paged block pool that our mechanism extends.
A line of work manages KV memory through virtual memory or custom allocators: vAttention~\cite{vattention}
uses UVM demand paging for contiguity, GMLake~\cite{gmlake} stitches fragmented allocations with VMM,
and vLLM ships a cumem-based sleep and wake path that unmaps whole allocations. Our contribution to
that line is a fine-grained, kernel-transparent two-handle-one-VA toggle. Jenga~\cite{jenga} and
related work size and place KV blocks within a fixed budget rather than changing the budget. Chunked
prefill, which our negative result turns on, is the scheduling technique that interleaves prefill
chunks with decode. Our finding is that it already resolves the prefill-versus-KV tension that
motivated an elastic budget in the first place.

\section{Conclusion}
We set out to reclaim the prefill-activation reserve in an LLM serving engine, built a userspace
elastic KV mechanism that toggles the reserve in milliseconds without touching the attention kernel
or the driver, and proved that a static commit of the reserve is unsafe so the dynamic toggle is
necessary to capture it. We then tested the premise the technique depends on and found it does not
hold: a small prefill chunk barely raises time-to-first-token, because prefill is compute bound and
decode is cheap per step, so lowering \code{max\_num\_batched\_tokens} recovers more KV at equal
latency than the elastic cache does. Tensor parallelism removes the only regime where the reserve is
large. The engineering works, but the opportunity does not, on the workloads that matter today. We
report this as a negative result because it is useful. It tells the community that vLLM's chunked prefill has
already closed the prefill-versus-KV-capacity gap, and it points at the narrow conditions under which
an elastic budget might still pay off. We release the elastic-VMM mechanism, which is a general
kernel-transparent way to grow and shrink a contiguous GPU tensor, in case it is useful elsewhere.

\paragraph{Reproducibility.}
All numbers are from a single A100-40GB running Qwen2.5-7B-Instruct on vLLM 0.23.0. The decisive TTFT
experiment, the necessity contradiction, and the toggle microbenchmarks are each driven by a small
standalone harness. The mechanism is a torch pluggable allocator plus a block-pool gate and a
scheduler-boundary controller, none of which modify the attention kernel.

{\small
\bibliographystyle{plainnat}
\bibliography{refs}
}
\end{document}